\documentclass[amsmath,superscriptaddress,prb,twocolumn] {revtex4}
\usepackage{bm}
\usepackage{enumerate}

\usepackage{graphicx}
\usepackage[dvips]{epsfig}
\usepackage{epsf}
\usepackage{xcolor}

\makeatletter
\newcommand{\Rmnum}[1]{\expandafter\@slowromancap\romannumeral #1@}
\makeatletter

\begin{document}
\title{Anomalous Hall effect in metallic collinear antiferromagnets with charge-density wave order}
\author{Vladimir A. Zyuzin}
\affiliation{L.D. Landau Institute for Theoretical Physics, 142432, Chernogolovka, Russia}
\begin{abstract}
In this paper we propose two theoretical models of a metallic N\'{e}el ordered antiferromagnet with charge-density wave order which shows the anomalous Hall effect. In our models a combination of the N\'{e}el order and the charge-density wave order results in spin splitting of the conducting fermions. In addition, spin-orbit coupling is required to drive the anomalous Hall effect. In the first model we analyzed the effects of the Rashba spin-orbit coupling, and of the intrinsic $d-$wave spin-orbit coupling in the second model. 
\end{abstract}
\maketitle

\section{Introduction}
There is a notion of weak ferromagnetism in insulating antiferromagnets when N\'{e}el order acquires a small magnetic moment due to the Dzyaloshinskii-Moriya interaction which is the the spin-orbit coupling interaction \cite{Dzyaloshinskii1958}.
It has been then proposed that spin-wave excitations about the N\'{e}el order can result in spin Nernst effect \cite{Cheng2016,ZyuzinKovalev2016} which is an indication of a non-zero spin-wave carried finite magnetization. It exists only at non-zero temperatures.
Recently it has been proposed that N\'{e}el ordered metallic antiferromagnets may also have finite magnetic moment due to orbital magnetization of the conducting fermions which interact with the N\'{e}el order through momentum dependent exchange interaction \cite{AHE_AFM,AHE_AFM_Review}. Such antiferromagnets are suggested to be called as the N\'{e}el ordered altermagnets \cite{AHE_AFM}. The N\'{e}el order is intact in this scenario which makes it different from the weak Dzyaloshinskii type ferromagnetism. We point out that the spin-orbit coupling is also required for the conducting fermions to interact with in order to show the magnetic moment and, therefore, show anomalous Hall effect. 
In this paper we show that there is another option of having weak ferromagnetism due to the orbital magnetic moment of conducting fermions in metallic N\'{e}el ordered antiferromagnets. It requires, in addition to the N\'{e}el order, a charge-density wave order which results in effective Zeeman type splitting of conducting fermions. Again, the spin-orbit coupling, which locks fermion's spin with the momentum, together with such effective Zeeman splitting results in the orbital magnetic moment of conducting fermions. Thus, such antiferromagnets are going to show anomalous Hall effect upon passing of the electric current through the system.
We discuss two types of spin-orbit coupling. In our first model it is the Rashba spin-orbit coupling and we map our results to the known Rashba and Zeeman model of anomalous Hall effect \cite{CulcerMacDonaldNiuPRB2003,AHE_RMP}.
In our second model we propose intrinsic $d-$wave spin-orbit coupling which generates anomalous Hall effect.

It is instructive to go over the details of anomalous Hall effect in the most simple N\'{e}el ordered antiferromagnets with two sublattices.
It us understood \cite{AHE_AFM,AHE_AFM_Review} that if a combination of time-reversal operation and translation which connects the two sublattices exists, then the anomalous Hall effect is absent in the N\'{e}el ordered antiferromagnets. In Refs. \onlinecite{AHE_AFM,AHE_AFM_Review} it was shown that the anomalous Hall effect can appear in N\'{e}el ordered antiferromagnets when the translation in the mentioned above combination is broken and a rotation operation together with the time-reversal operation is the only symmetry which connects the two sublattices of the order. Here we show that the anomalous Hall effect will also exist in N\'{e}el ordered antiferromagnets when the translation as well as the rotation are broken in the combination. In this case the time-reversal symmetry is explicitly broken and the system is a weak ferromagnet due to the spin-splitting of the conducting fermions.

\section{Model A}
\begin{figure}[t] 
\includegraphics[width=0.25 \columnwidth ]{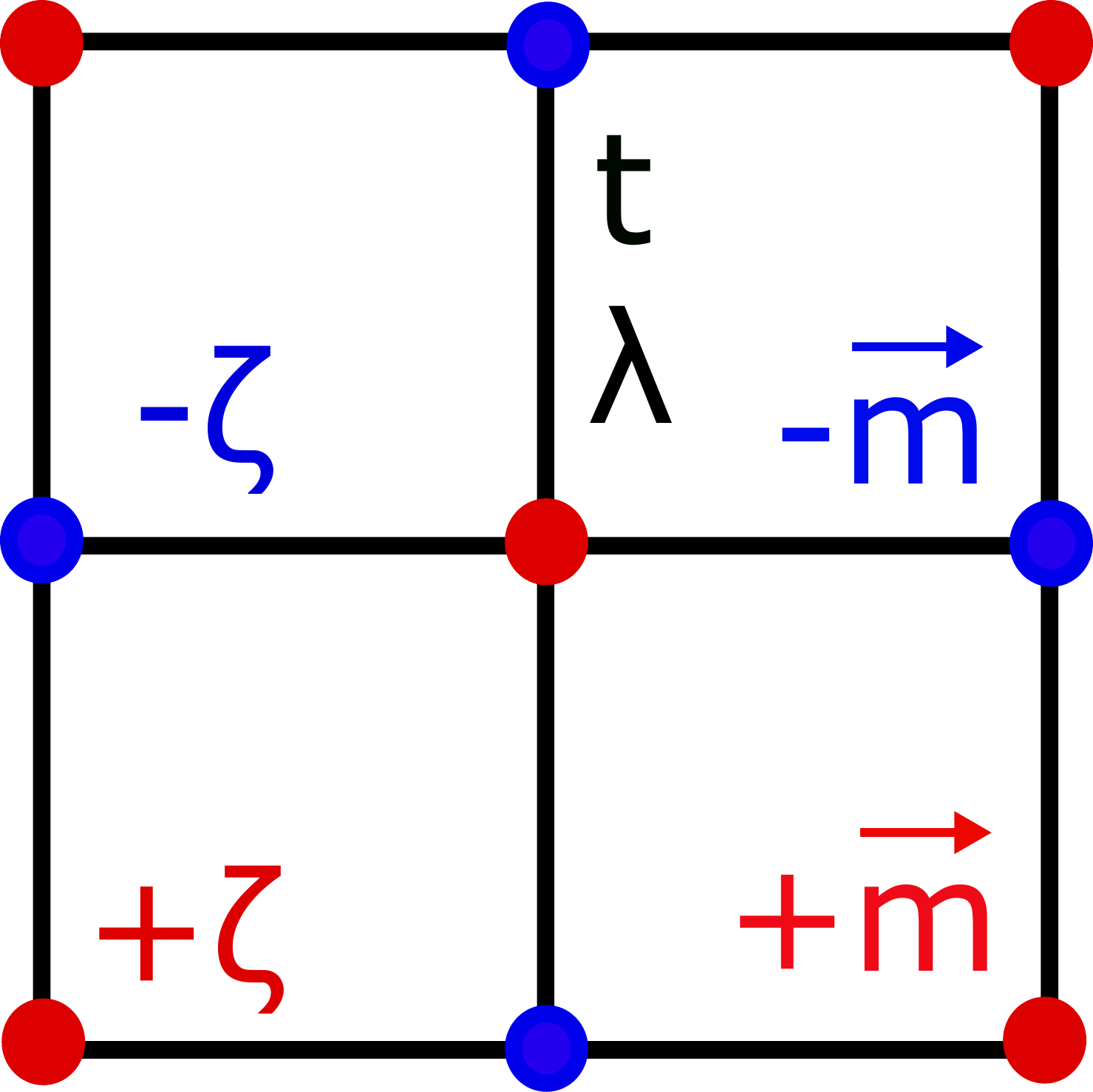} ~~~~~~~~~~
\includegraphics[width=0.292 \columnwidth]{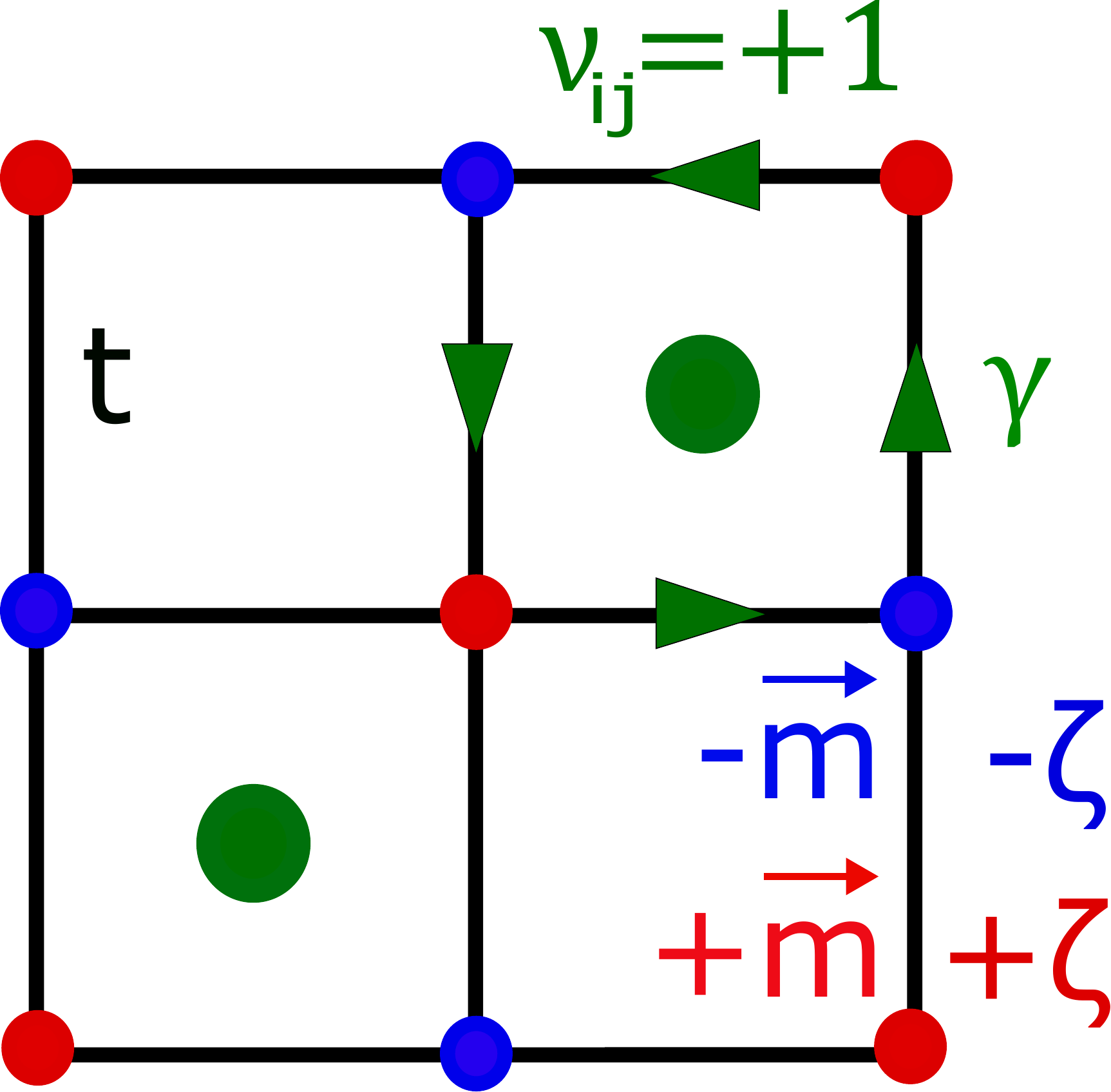}

\protect\caption{Left: two-dimensional square lattice. There are two sublattices marked by red and blue cites. Red cites have localized magnetic moments pointing in $+{\bf m} = (m_{x},m_{y},m_{z})$ direction, while blue ones have localized magnetic moments in $-{\bf m}$ direction. This is the N\`{e}el order. We study it within the mean-field in a sense that ${\bf m}$ is fixed and doesn't fluctuate.
There is a charge-density wave order which is characterized by different electric potential on the two sublattices the conducting fermions interact with, red cites have $+\zeta$ while blue cites have $-\zeta$.
Nearest neighbor hopping is $t$ and there is a Rashba spin orbit coupling $\lambda$.
Right: all objects and notations are the same as in the left figure, except for the
green circles which correspond to the non-magnetic atoms which result in the out of plane spin-orbit coupling denoted by $\gamma$ and by directions of the green arrows for which $\nu_{ij} = +1$. In addition, these atoms result in anisotropic second-nearest neighbor hopping of conducting fermions. Indeed, hopping across plaquette with the green atom may be different than across the plaquette without.}
\label{fig:fig1}  
\end{figure}

\begin{figure}[h] 
\includegraphics[width=0.48 \columnwidth]{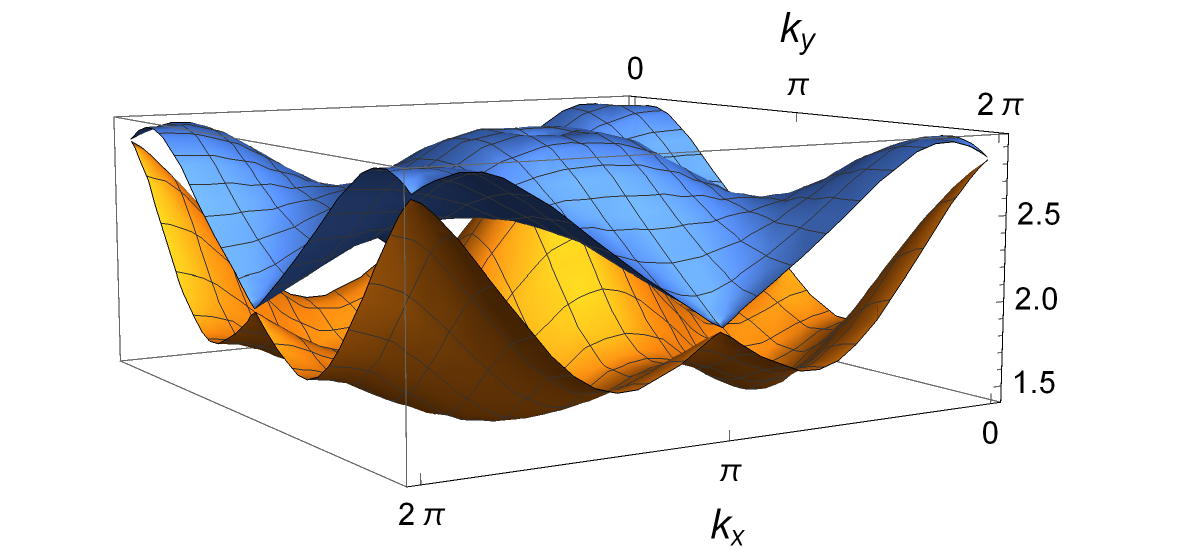} 
\includegraphics[width=0.48 \columnwidth]{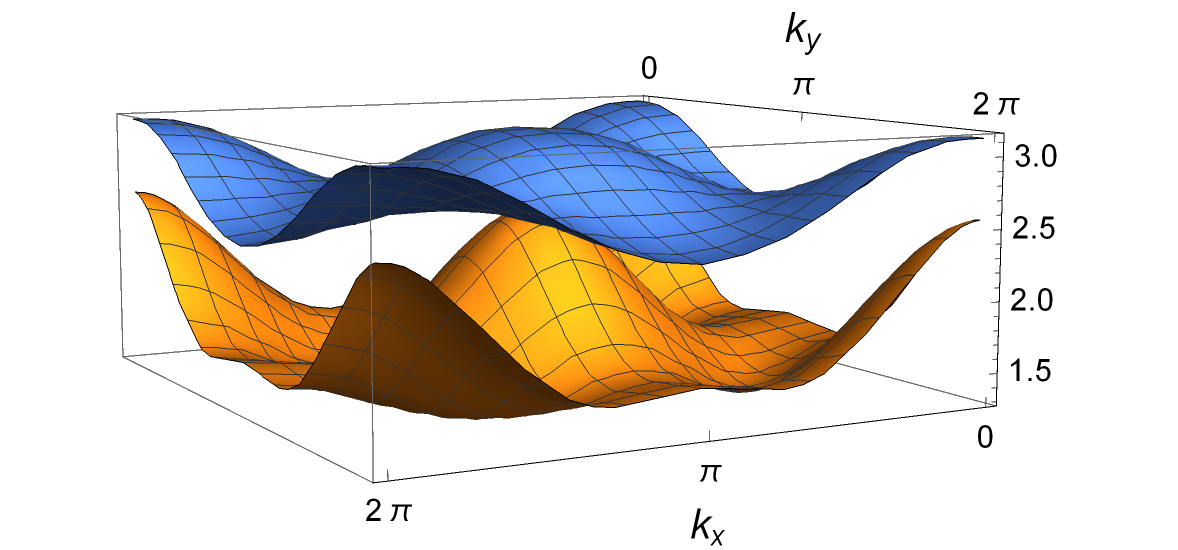}

\protect\caption{Plot of the $\epsilon_{{\bf k};\pm;+}$ spectrum of the Eq. (\ref{modelA}) for $t=1$, $\lambda = 0.4$, $m_{z}=2$ and $m_{x}=m_{y}= 0$ parameters in units of $t$. $\epsilon_{{\bf k};+;+}$ is in blue, while $\epsilon_{{\bf k};-;+}$ is in yellow. Left:  $\zeta = 0$ and right $\zeta = 0.4$. From left to right: charge-density wave order in combination with the antiferromagnetism is shown to open up a gap at the degeneracy point of the two spectra.
}
\label{fig:fig2}  
\end{figure}

\begin{figure}[h] 
\centerline{
\includegraphics[width=0.45 \columnwidth , trim=0 0cm 0 0 ]{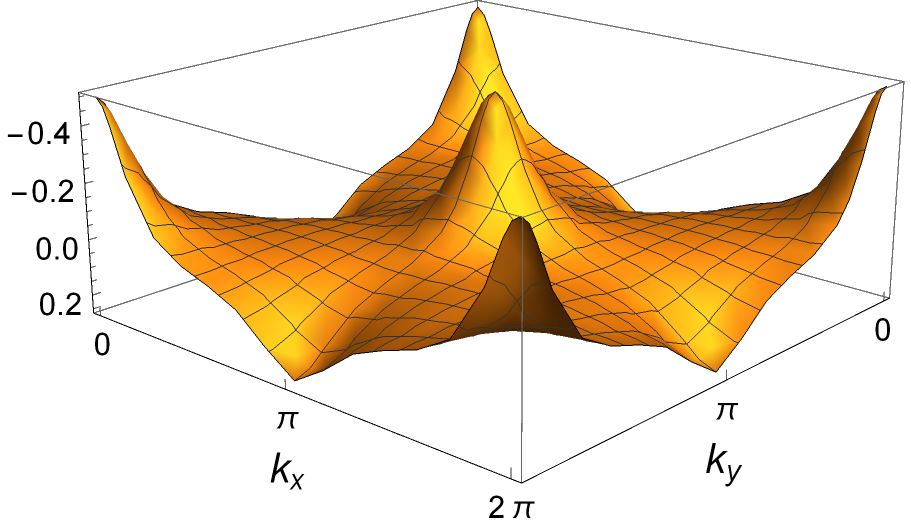} ~
\includegraphics[width=0.45 \columnwidth]{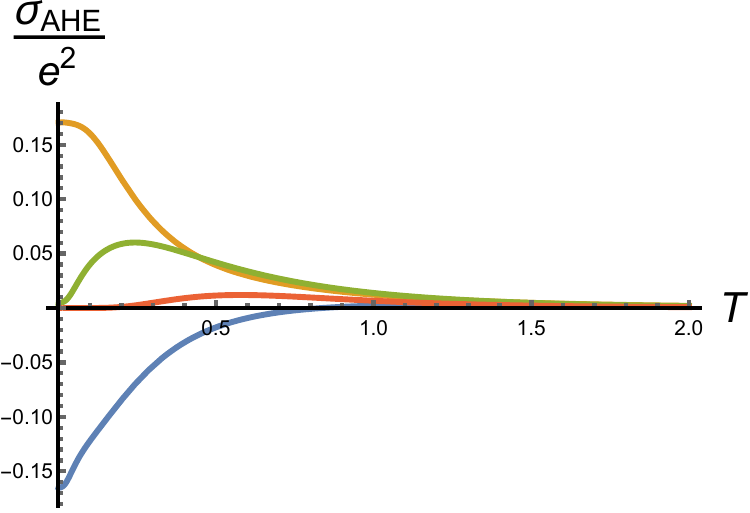}
}
\protect\caption{Left: plot of the Berry curvature $\Omega_{{\bf k};\pm;+}$ for $\epsilon_{{\bf k};+;+}$ band of the model Eq. (\ref{modelA}) for $t=1$, $\lambda = 0.4$, $m_{z}=2$, $m_{x}=m_{y}= 0$, and $\zeta=0.4$ parameters in units of $t$. 
Berry curvatures corresponding to other energy bands are obtained from $\Omega_{{\bf k};\pm;s} = \Omega_{{\bf k};\pm;-s}$ and  $\Omega_{{\bf k};+;s} = - \Omega_{{\bf k};-;s}$ relations.
 Right: Plot of the anomalous Hall conductivity as a function of temperature for blue $\mu =3$, yellow $\mu=2$, green $\mu=1.4$, red $\mu=0.5$. The parameters are chosen to be  $t=1$, $\lambda = 0.4$, $\zeta = 0.4$, $m_{z}=2$ and $m_{x}=m_{y}= 0$ in units of $t$. We have set $h =2\pi \hbar \equiv 1$.
}
\label{fig:fig3}  
\end{figure}

We consider a two-dimensional square lattice with two sublattices corresponding to the N\'{e}el order. There are conducting fermions that interact with the N\'{e}el order through an exchange interaction, which has opposite signs on the two sublattices according to the structure of the N\'{e}el order. Please see left figure in Fig. (\ref{fig:fig1}) for details. Furthermore, there is a different electric potential on the two sublattices corresponding to the charge-density wave order. The charge-density wave order breaks any discussed above symmetry which connects the two sublattices.
In addition, we add Rashba spin-orbit coupling needed to couple momentum with spin. Such spin-orbit coupling may occur naturally at the two-dimensional surface of a three-dimensional material, in polar layered systems, and in many more situations.
We introduce a tight-binding Hamiltonian corresponding to the described above system,
\begin{align}\label{modelA}
\hat{H}_{\mathrm{A}} = 
\left[
\begin{array}{cc} 
\zeta + {\bf m}\cdot{\bm \sigma}  & \xi_{\bf k} + \lambda\left[ s_{x}\sigma_{y} - s_{y}\sigma_{x} \right]  \\
 \xi_{\bf k} + \lambda\left[ s_{x}\sigma_{y} - s_{y}\sigma_{x} \right] & -\zeta - {\bf m}\cdot{\bm \sigma} 
 \end{array}
 \right],
\end{align}
where $\xi_{\bf k} = -t [\cos(k_{x}) + \cos(k_{y}) ]$ is the nearest neighbor hopping, $s_{x/y} \equiv \sin(k_{x/y})$ notation is used for brevity, $\zeta$ is a parameter responsible for the charge-density wave order and ${\bf m} = (m_{x},m_{y},m_{z})$ is the N\'{e}el order, parameter $\lambda$ is the Rashba spin-orbit coupling. The Hamiltonian acts in the space of sublattices and spin, such that the overall spinor structure is $\Psi^{\dag} = (\psi^{\dag}_{{\mathrm{R},\uparrow}},\psi^{\dag}_{{\mathrm{R},\downarrow}},\psi^{\dag}_{{\mathrm{B},\uparrow}},\psi^{\dag}_{{\mathrm{B},\downarrow}})$, where $\mathrm{R}$ and $\mathrm{B}$ stand for red and blue correspondingly and arrows stand for spin. Pauli matrices ${\bm \sigma}$ act on spin of fermions. We study the N\'{e}el order as well as the charge-density wave order within the mean-field in a sense that ${\bf m}$ and $\zeta$ are fixed and don't fluctuate.
Spectrum of the system is 
\begin{align}
\epsilon^2_{{\bf k};\pm} & = m^2+ \zeta^2  + \xi_{\bf k}^2 + \lambda^2 (s_{x}^2 + s_{y}^2) 
\label{spectrum}
\\
&
\pm
\bigg[
4\zeta^2 m^2  -
8\lambda\zeta\xi_{\bf k}(m_{x}s_{y}-m_{y}s_{x})
\nonumber
\\ 
&
+
4\lambda^2(m_{x}s_{y}-m_{y}s_{x})^2
+ 4\lambda^2  (s_{x}^2 + s_{y}^2) (m^2 + \xi_{\bf k}^2 )
\bigg]^{\frac{1}{2}}
\nonumber
\end{align}
Let us first focus on the case when $m_{z}\neq 0$ and $m_{x}=m_{y}=0$. The spectrum is more simpler in this case,
\begin{align}
\epsilon^2_{{\bf k};\pm} & =m^2+ \zeta^2 + \xi_{\bf k}^2 + \lambda^2 (s_{x}^2 + s_{y}^2) 
\nonumber
\\
&
\pm
\sqrt{
4\zeta^2 m^2 + 4\lambda^2  (s_{x}^2 + s_{y}^2) (m^2 + \xi_{\bf k}^2 )
}.
\end{align}
The spectrum $\epsilon_{{\bf k};\pm;s=\pm} = s \epsilon_{{\bf k};\pm}$ has an energy gap $2\vert m \mp \zeta \vert$ due to the antiferromagnetic ${\bf m}\neq 0$ and charge-density wave $\zeta \neq 0$ orders.
In what follows we assume that the valence bands $\epsilon_{{\bf k};\pm;-}$ are always occupied, which can be achieved with the choice of the Fermi level being either in the energy gap or in the conduction band $\epsilon_{{\bf k};\pm;+}$. 
We plot $\epsilon_{{\bf k};\pm;+}$ in Fig. (\ref{fig:fig2}), left for the case when $\zeta = 0$ and right for $\zeta = 0.4$ in units of $t$. 
The spectrum in the right plot in Fig. (\ref{fig:fig2}) can be described by the Rashba and Zeeman model which is known to show anomalous Hall effect \cite{CulcerMacDonaldNiuPRB2003,AHE_RMP}. 
We can understand effective Zeeman splitting by squaring the Hamiltonian Eq. (\ref{modelA}) in which we set $\lambda=0$ for simplicity of the argument,
\begin{align}
\hat{H}_{\mathrm{A}}^2 = \zeta^2 + m^2 +\xi_{\bf k}^2 + 2\zeta{\bf m}\cdot{\bm \sigma}.
\end{align} 
The last term is exactly the effective Zeeman type splitting of conducting fermions achieved by the interplay of N\'{e}el and charge-density wave orders. Upon addition of Rashba spin-orbit coupling, $\lambda\neq 0$, fermions will acquire the Berry curvature, which we plot in Fig. (\ref{fig:fig3}). 
Intrinsic mechanism of the anomalous Hall effect \cite{AHE_RMP} is given by an integral of the product of the Berry curvature and the distribution function over the Brillouin zone,
\begin{align}\label{currentAHE}
{\bf j}_{\mathrm{AHE}} & = e^2
\left[ \int_{\mathrm{BZ}}\frac{d {\bf k}}{(2\pi)^2}\sum_{s=\pm;n = \pm} {\bm \Omega}_{{\bf k};n;s} 
{\cal F}(\epsilon_{{\bf k};n;s} )\right]\times {\bf E} 
\\
&
\equiv \sigma_{\mathrm{AHE}} \mathrm{sign}(m_{z}\zeta) {\bf e}_{z}\times{\bf E},
\nonumber
\end{align}
where ${\bm \Omega}_{{\bf k};n;s} = \Omega_{{\bf k};n;s}{\bf e}_{z}$ is the Berry curvature of the $\epsilon_{{\bf k};n;s}$ band plotted in Fig. (\ref{fig:fig3}), ${\cal F}(\epsilon) = (e^{\frac{\epsilon - \mu}{T}}+1 )^{-1}$ is the Fermi-Dirac distribution function, where $T$ is the temperature and $\mu$ is the Fermi level. We find that $\Omega_{{\bf k};\pm;s} = \Omega_{{\bf k};\pm;-s}$ and  $\Omega_{{\bf k};+;s} = - \Omega_{{\bf k};-;s}$ in the studied model.
We plot the anomalous Hall effect conductivity in right plot of Fig. (\ref{fig:fig1}) for different values of $\mu$. 
The anomalous Hall conductivity vanishes at exactly $\mu=0$, i.e. when the Fermi level is in the middle of the energy gap, for any temperatures. 
When $\mu \neq 0$ but is still in the energy gap, the anomalous Hall conductivity vanishes only at $T=0$, but might be finite at $T>0$ (see red plot in the right figure of Fig. (\ref{fig:fig3}). 
Only the $\epsilon_{{\bf k};-;+}$ out of the two conduction bands is occupied when $\mu=2$ and $\mu=1.4$ in units of $t$. When $\mu=3$ both $\epsilon_{{\bf k};\pm;+}$ are occupied. In Fig. (\ref{fig:fig6}) we plot anomalous Hall conductivity for other parameters corresponding to a small value of the N\'{e}el order order parameter, and where as a result there is no energy gap in the spectrum of fermions.
We note that the sign of anomalous Hall effect conductivity in our model depends on the sign of the $m_{z}\zeta$ product.


Now let us briefly mention the case of the in-plane N\'{e}el order, namely $m_{x}\neq 0$, $m_{y}\neq 0$ and $m_{z} = 0$. From Eq. (\ref{spectrum}) we observe that there is still an energy gap $2\vert m \mp \zeta \vert$ in the spectrum. However, the Berry curvature is zero in this case and there is no anomalous Hall effect. From here we conclude that anomalous Hall effect in our model depends on $m_{z}$ direction of the N\'{e}el order and $\zeta$ only.

\section{Model B}
Let us now introduce another example of the N\'{e}el ordered metallic antiferromagnet with charge-density wave order but without Rashba spin-orbit coupling. We study fermions on a square lattice depicted in the right figure of Fig. (\ref{fig:fig1}) described by the following Hamiltonian,
\begin{align}\label{modelB}
\hat{H}_{\mathrm{B}} = 
\left[
\begin{array}{cc} 
\delta_{\bf k} + {\bf m}\cdot{\bm \sigma} & \xi_{\bf k} + i\gamma_{\bf k}\sigma_{z} \\
 \xi_{\bf k} - i\gamma_{\bf k}\sigma_{z} & -\delta_{\bf k} - {\bf m}\cdot{\bm \sigma}
 \end{array}
 \right],
\end{align}
where $\delta_{\bf k} = \zeta + \omega_{\bf k}$, where $\zeta$ corresponds to the charge-density wave order and $\omega_{\bf k} = \omega \sin(k_{x})\sin(k_{y})$ is the anisotropic second nearest neighbor hopping. Such an anisotropy is achieved with the green atom in the right figure of Fig. (\ref{fig:fig1}). We assume that there is no fermion state on the green atom. Then, hopping across empty plaquette is different from the one with a green atom. In addition to $\omega_{\bf k}$ the green atoms result in intrinsic $d-$wave spin-orbit coupling given by $\gamma_{\bf k } = \gamma \left[\cos(k_{x}) - \cos(k_{y})\right]$.
\begin{figure}[h] 
\centerline{
\includegraphics[width=0.45 \columnwidth, trim=0 0.5cm 0 0]{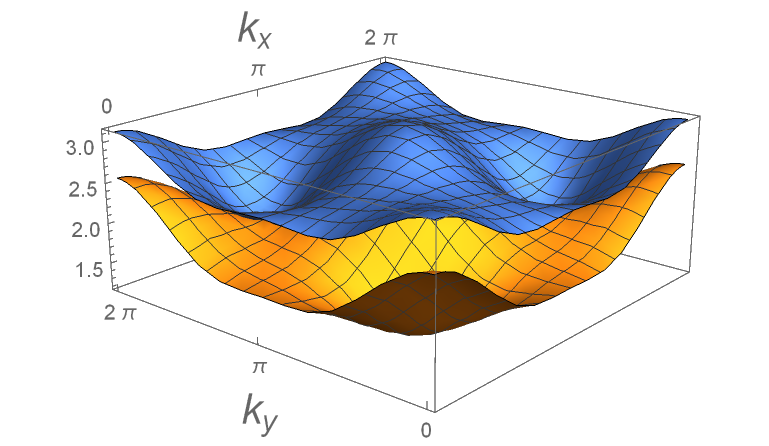}~~
\includegraphics[width=0.45 \columnwidth]{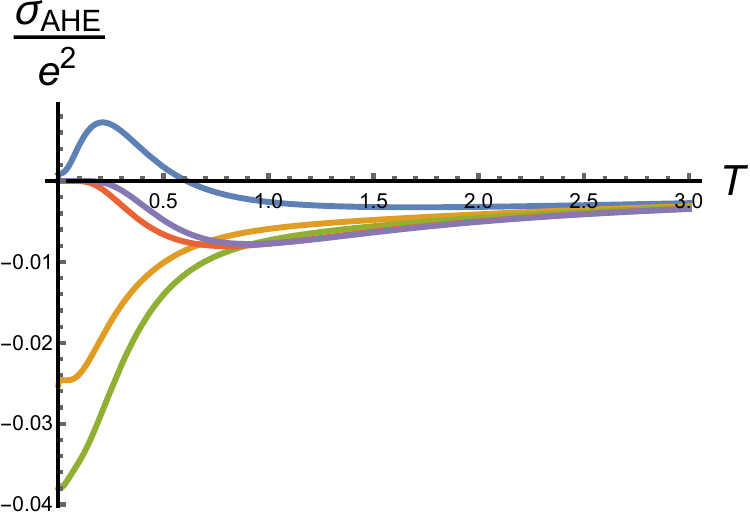}
}
\protect\caption{Left: spectrum $\epsilon_{{\bf k};\pm;+}$ defined in Eq. (\ref{spectrumB}) of the model Eq. (\ref{modelB}). There are two other branches of the spectrum given by $\epsilon_{{\bf k};\pm;-} = -\epsilon_{{\bf k};\pm;+} $. Right: plot of the anomalous Hall conductivity as a function of temperature for blue $\mu =3$, yellow $\mu=2$, green $\mu=1.4$, red $\mu=0.5$, purple $\mu=0$. 
The parameters on both figures are chosen to be $t=1$, $m_{z}=\sqrt{3.5}$, $m_{x}=\sqrt{0.5}$, $m_{y}=0$, $\gamma = 0.4$, $\omega=0.4$, and $\zeta=0.4$ parameters in units of $t$. We have set $h =2\pi \hbar \equiv 1$.
}
\label{fig:fig4}  
\end{figure}
 
Spectrum of fermions is
\begin{align}\label{spectrumB}
\epsilon^2_{{\bf k};\pm} =  m^2 + \xi_{\bf k}^2 + \delta_{\bf k}^2 + \gamma_{\bf k}^2 \pm 2\sqrt{m^2\delta_{\bf k}^2 + m_{\parallel}^2\gamma_{\bf k}^2}, 
\end{align}
which is plotted in Fig. (\ref{fig:fig4}), left. The structure of the spectrum can be understood from
\begin{align}\label{modelBsq}
\hat{H}^2_{\mathrm{B}} &= m^2 + \xi_{\bf k}^2 + \delta_{\bf k}^2 + \gamma_{\bf k}^2
\nonumber
\\
& + 2{\bf m}\cdot{\bm \sigma} \delta_{\bf k}\tau_{0} + [{\bf m}\cdot{\bm \sigma},i\gamma_{\bf k}\sigma_{z}]\tau_{1}, 
\end{align}
where $\tau_{0}$ and $\tau_{1}$ are the Pauli matrices acting in the sublattice space. First term in the second line of Eq. (\ref{modelBsq}) is the spin-splitting of conducting fermions, $\propto 2{\bf m}\cdot{\bm \sigma}\delta_{\bf k}\tau_{0}$, in which $\zeta$ is the Zeeman like spin-splitting discussed above and $\omega_{\bf k}$ is the $d-$wave spin-splitting \cite{AHE_AFM} (it was suggested to call such momentum-dependent spin splitting as the altermagnetism \cite{SmejkalSinovaJungwirth2022a,SmejkalSinovaJungwirth2022b}).  It is interesting to note that such a term was proposed in Refs. \onlinecite{VarmaZhu2006,WuSunFradkinZhang2007} to appear spontaneously as a result of Pomeranchuk instability in higher harmonics channel. Here it appears as a byproduct of the antiferromagnetic instability which leads to non-zero ${\bf m}$ provided that there is an anisotropic second-nearest neighbor fermion hopping achieved by the green atoms in the right figure of Fig. (\ref{fig:fig1}). Upon addition of in-plane magnetic field and Rashba spin-orbit coupling, this model will show the $d-$wave Hall effect \cite{VorobevZyuzin}. A three-dimensional generalization of the model with addition of an appropriate spin-orbit coupling provided that the N\'{e}el order is set to the $x-y$ plane will show anomalous Hall effect \cite{AHE_AFM,AHE_AFM_Review,SmejkalSinovaJungwirth2022a,SmejkalSinovaJungwirth2022b}. 
\begin{figure}[h] 
\centerline{
\includegraphics[width=0.45 \columnwidth , trim=0 0cm 0 0 ]{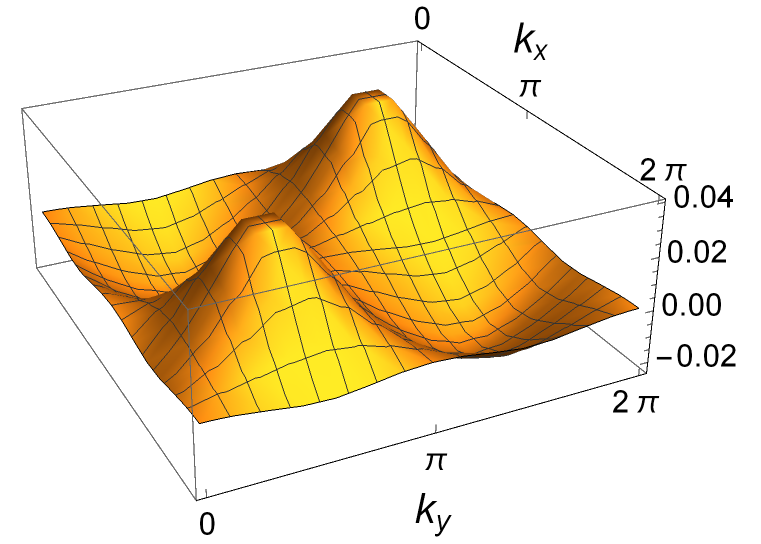} ~
\includegraphics[width=0.45 \columnwidth]{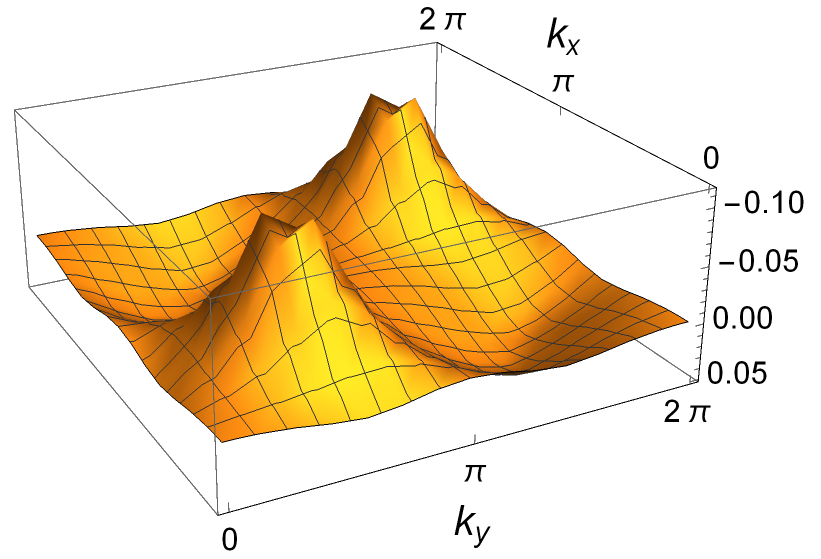}
}
\protect\caption{Plot of the Berry curvature $\Omega_{{\bf k};\pm;+}$ for $\epsilon_{{\bf k};\pm;+}$ bands (left for + and right for -) of the model Eq. (\ref{modelB}) for $t=1$, $m_{z}=\sqrt{3.5}$, $m_{x}=\sqrt{0.5}$, $m_{y}=0$, $\gamma = 0.4$, $\omega=0.4$, and $\zeta=0.4$ parameters in units of $t$. 
Berry curvature of the $\epsilon_{{\bf k};\pm;-}$ bands are that of the $\epsilon_{{\bf k};\pm;+}$ bands but with opposite sign, i.e. $\Omega_{{\bf k};\pm;-} = - \Omega_{{\bf k};\pm;+}$.
}
\label{fig:fig5}  
\end{figure}

Our model Eq. (\ref{modelB}) introduces another possibility of anomalous Hall effects in collinear antiferromagnets by virtue of the intrinsic $d-$wave spin-orbit coupling, second term in the second line of Eq. (\ref{modelBsq}), which is non-zero when the N\'{e}el order has a component in the $x-y$ plane. Indeed, we can find a configuration of the order, in which all three spin matrices are involved in the Hamiltonian describing fermions.
For example, for $m_{y}=0$ in the second line of Eq. (\ref{modelBsq}) we have $2m_{z}\zeta\tau_{0}\sigma_{z}$, $2m_{x}\omega_{\bf k}\tau_{0}\sigma_{x}$ and $2m_{x}\gamma_{\bf k}\tau_{1}\sigma_{y}$ terms. These are the required terms for the anomalous Hall effect to occur. There is also a $2m_{z}\omega_{\bf k}\tau_{0}\sigma_{z}$ term but it doesn't lead to the anomalous Hall effect as its momentum dependence is the same as in the term with $\sigma_{x}$ spin matrix.

We plot the Berry curvature for different bands in Fig. (\ref{fig:fig5}) and observe that unlike in the model Eq. (\ref{modelA}) Berry curvatures of the fermion energy bands of the model Eq. (\ref{modelB}) are not equal in magnitude. 
Namely, $\Omega_{{\bf k};\pm;s} = - \Omega_{{\bf k};\pm;-s}$ and  $\vert \Omega_{{\bf k};+;s} \vert \neq  \vert \Omega_{{\bf k};-;s}\vert$.
As a result, there is a non-zero anomalous Hall effect at $T\neq 0$ when chemical potential is set to zero, i.e. the system is an insulating antiferromagnet. The anomalous Hall effect is plotted in the right of Fig. (\ref{fig:fig4}). Purple curve corresponds to $\mu=0$, in which case the anomalous Hall effect is absent at $T=0$ being a consequence of the cancellation of the integrated Berry curvatures over the Brillouin zone of the occupied bands. At $T\neq 0$ the cancellation no longer holds and in addition unoccupied bands also start contributing to the anomalous Hall conductivity.  It is worth mentioning that $\mu=0$ is not a Chern insulator and there are no edge states in the $2\vert m \mp \zeta \vert$ energy gap due to the antiferromagnetic and charge-density wave orders. This is because non-trivial topology is achieved indpendently in the conduction and valence bands, and we rather expect edge states in the conduction and valence energy bands.
We note, there is a proposal of obtaining a quantum anomalous Hall effect \cite{Guo_npj2023} in a system similar to Eq. (\ref{modelB}) but with rather exotic structure of the second-nearest neighbor hopping. 

We note that the magnitude of the anomalous Hall conductivity is rather small at large temperatures, but it is non-zero. This is another consequence of inequivalent magnitudes of Berry curvatures of different energy bands. As seen in the right plot of Fig. (\ref{fig:fig3}) the anomalous Hall effect in the model Eq. (\ref{modelA}) vanishes at large temperatures because of the equal in magnitude Berry curvatures of different energy bands.
The anomalous Hall conductivity in the model Eq. (\ref{modelB}) is $\propto \mathrm{sign}(m_{z}\zeta \omega \gamma)$, is zero if either of the four components is zero, and is unexpectedly $\propto m_{\parallel}^2$ rather than to a $d-$wave combination of $m_{x}$ and $m_{y}$.
Since changing sign of $\omega$ will automatically change sign of $\gamma$, the sign dependence of the anomalous Hall conductivity in the model Eq. (\ref{modelB}) can be reduced to a $\propto \mathrm{sign}(m_{z}\zeta )$, exactly as in the model Eq. (\ref{modelA}). The sign change of the $\sigma_{\mathrm{AHE}}$ as a function of Fermi level $\mu$ or temperature $T$ shown in the right figures of Fig. (\ref{fig:fig3}) and (\ref{fig:fig4}) is because Berry curvatures of different conduction bands $\epsilon_{{\bf k};\pm;+}$ are opposite in sign, as well as the Berry curvature itself is sign changing as a function of momentum (as shown in Fig. (\ref{fig:fig3}) and (\ref{fig:fig5})).

\section{Discussion}
We now wish to connect our proposed models to cuprates, where polar Kerr effect has been observed \cite{cuprateEXP}. 
First of all, polar Kerr effect occurs in the pseudogap phase of cuprates where charge-density wave order exists and the system is close to the antiferromagnetic state. 
Furthermore, the Kerr signal has the same sign from the opposite surfaces and it is known that there is no Faraday rotation. 
Moreover, the Kerr effect can't be trained by external magnetic field.
Finally, the magnitude of the observed Kerr rotation is four orders less than that from other itinerant ferromagnetic oxides \cite{cuprateEXP}, suggesting that only a very weak magnetic moment exists in the pseudogap phase of cuprates.

\begin{figure}[h] 
\centerline{
\includegraphics[width=0.45 \columnwidth, trim=0 -0cm 0 0]{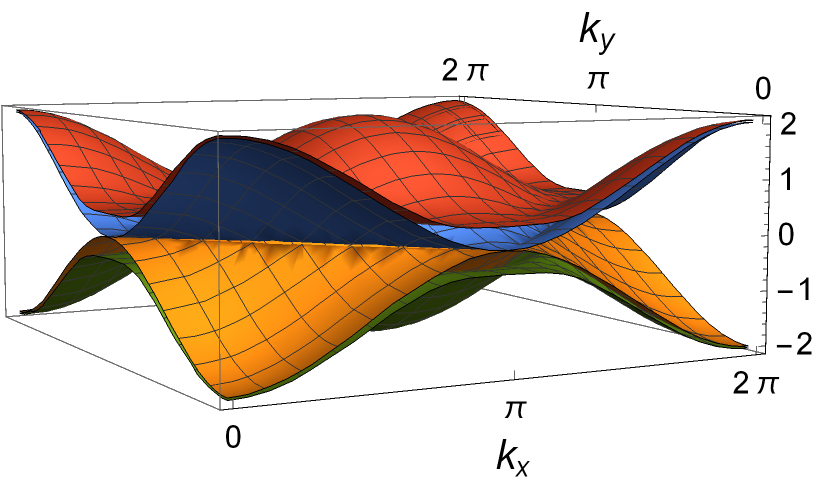}~~
\includegraphics[width=0.45 \columnwidth]{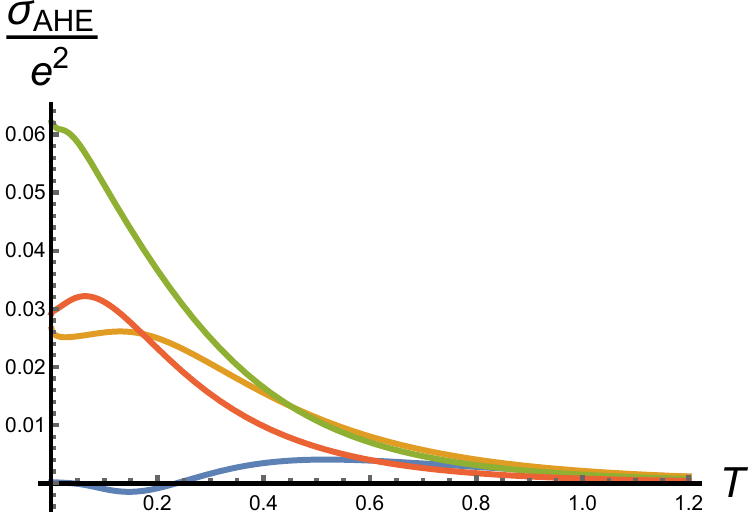}
}
\protect\caption{
Left: spectrum of the model Eq. (\ref{modelA}). Right: plot of the anomalous Hall conductivity of the model Eq. (\ref{modelA}) as a function of temperature for blue $\mu =1.5$, yellow $\mu=1$, green $\mu=0.5$, and red $\mu=0.25$ in units of $t$. 
The parameters on both figures are chosen to be $t=1$, $m_{z}=0.2$, $m_{x}=m_{y}=0$, $\zeta=0.2$, $\lambda=0.05$ parameters in units of $t$. We have set $h =2\pi \hbar \equiv 1$.
}
\label{fig:fig6}  
\end{figure}

Polar Kerr effect and Faraday effect are defined by the imaginary and real parts of $\sigma_{\mathrm{AHE}}(\omega)$ correspondingly \cite{Kapitulnik2015} (please also see \cite{PershogubaKechedzhiYakovenko2013, Hosur2013, Mineev2013}). 
Therefore, we expect them to occur in both of our models Eq. (\ref{modelA}) and Eq. (\ref{modelB}). 
One can create a three-dimensional structure out of the studied models Eq. (\ref{modelA}) and Eq. (\ref{modelB}) by stacking them in the $z-$direction. 
In the case of the model Eq. (\ref{modelA}) Rashba spin-orbit coupling will then exist only in the surface layers, and the anomalous Hall effect will happen only in the surface layers. 
Then there will be no Faraday effect because anomalous Hall effect will vanish in the bulk, but there will be Kerr effect upon reflection of electromagnetic wave from the surfaces. 
Same sign of polar Kerr effect from opposite surfaces can be achieved in our model by doubling the unit cell in $z-$direction and
making the two layers in the unit cell different only in the sign of either ${\bf m}$ or $\zeta$, for example, by layering $(+{\bf m},+\zeta)$ and $(+{\bf m},-\zeta)$. 
Or quadrupling the unit cell by, for example, layering $(+{\bf m},+\zeta)$, $(+{\bf m},-\zeta)$, $(-{\bf m},-\zeta)$, and $(-{\bf m},+\zeta)$. 
Absence of training by external magnetic field can be explained by inability to flip the N\'{e}el order by the magnetic field.
The advantage of model Eq. (\ref{modelB}) is that the anomalous Hall effect in model Eq. (\ref{modelB}) does not vanish at large temperatures.
Discussed above doubling or quadrupling of the unit cell of the model Eq. (\ref{modelB}) will also result in zero Faraday rotation in the bulk and same in sign polar Kerr effect from opposite surfaces.
We finally note, that in both models the magnitude of the anomalous Hall effect is small, please see Figs. (\ref{fig:fig3}) and (\ref{fig:fig4}), as compared to conductivity quantum (which would be $1$ in Fig. (\ref{fig:fig3}) and (\ref{fig:fig4})). 
For $t\approx 300 ~\mathrm{meV}$, the temperature at which charge-density wave appears is $T\approx 0.05t$. As shown in Fig. (\ref{fig:fig3}) and (\ref{fig:fig4}) this corresponds to the maximum values of $\sigma_{\mathrm{AHE}}$ for the cases when the Fermi level is in the conduction band (or in the valence band).  
If we take $\sigma_{\mathrm{AHE}}/e^2 \sim 1$ from a model of AHE in a ferromagnet \cite{CulcerMacDonaldNiuPRB2003}, then, according to Fig. (\ref{fig:fig3}) and (\ref{fig:fig4}), AHE in our models is $1-3$ orders of magnitude less than that in a ferromagnet. This is qualitatively consistent with the experimental observations of \cite{cuprateEXP}.
We have checked that such smallness is also the case for other values of parameters (for example please see Fig. (\ref{fig:fig6})).

There are other models which describe polar Kerr effect in cuprates \cite{Yakovenko2016,WangChubukovNandkishore2014}. 
These models don't rely on the Berry curvature due to the spin of the conducting fermions.

The other candidates for which our models Eq. (\ref{modelA}) and (\ref{modelB}) of AHE can be relevant to are \textit{R}NiC$_2$ (R =Gd,Tb) \cite{RNiC2exp2016} and NdNiO$_2$ nickelates \cite{nickelates2022} where charge-density wave order and antiferromagnetism have been experimentally observed.

We stress that we have proposed only a toy model with two coexisting orders, antiferromagnet and charge-density wave, which exhibits the anomalous Hall effect. 
However, in reality it might be the case that there are fluctuations instead.
We think that a hybrid interaction consisting of a mixture of antiferromagnetic and charge-density wave fluctuations might be relevant in the pseudogap phase of cuprates in this case. We speculate that one has to introduce an order parameter which is a combination of both N\'{e}el and charge-density wave order parameters which separately are zero on average but their multiplication is not. In this way, there will be a spontaneous magnetic moment of the conducting fermions.

In Ref. \onlinecite{ZyuzinZyuzin2015} it was proposed that there will be chiral electromagnetic waves (later called as the chiral Berry plasmons by the others) propagating at the boundaries between domains with opposite $\sigma_{\mathrm{AHE}}$ (opposite magnetizations) as well as the boundary of the sample. These waves are the photon analogs of the quantum Hall edges first proposed in Ref. \onlinecite{ZhukovRaikh}. Conditions for existence of such waves should also be present in antiferromagnets with charge-density wave order when either charge-density wave or the N\'{e}el order split into domains. Finally, we note that our model Eq. (\ref{modelA}) will exhibit Majorana zero modes when $s-$wave superconductivity \cite{SatoTakahashiFujimoto} will be included in the case when only $\epsilon_{{\bf k};-}$ band is occupied.

\section{Conclusions}
To conclude, we have proposed a mechanism of Zeeman-like spin-splitting of conducting fermions in an N\'{e}el ordered antiferromagnet in case in addition a charge-density wave order is present in the system. The N\'{e}el order creates exchange interaction with opposite sign on the two sublattices the conducting fermions experience, while charge-density wave order creates different chemical potential for the conducting fermions on the two sublattices. The mechanism of Zeeman-like spin-splitting of conducting fermions is based on the interplay of the two orders. As a result of the spin-splitting, the anomalous Hall effect is expected in a system where both orders and certain spin-orbit coupling are present. Such systems may be thought of as weak ferromagnets due to the magnetic moment of the conducting fermions.
We have come up with two theoretical models (Eq. (\ref{modelA}) and Eq. (\ref{modelB})) which show such spin-splitting and calculated anomalous Hall conductivity in them. We think that our models might be relevant to the experiment which observed polar Kerr effect in the pseudogap phase of cuprates \cite{cuprateEXP}. The pseudogap phase is known to have charge-density wave order and strong antiferromagnetic fluctuations, and it has been claimed that the polar Kerr effect appears at temperatures at which pseudogap phase sets in \cite{cuprateEXP}.

\section{Acknowledgements} The author thanks A.M. Finkel'stein, M.M. Glazov and J. Sinova for helpful discussions.
The author is grateful to Pirinem School of Theoretical Physics. This work is supported by FFWR-2024-0016.

\end{document}